\documentclass[twocolumn, prl]{revtex4}
\pdfoutput=1
\usepackage{graphicx}
\usepackage{amsmath}
\usepackage{epsfig}
\begin{document}
\title{Shaping topological properties of the band structures in a shaken optical lattice  }
\author{ Shao-Liang Zhang, Qi Zhou}
\affiliation{ Department of Physics, The Chinese University of Hong Kong, Shatin, New Territories, HK}
\date{\today}
\begin{abstract}
To realize band structures with non-trivial topological properties in an optical lattice  is an exciting topic in current studies on ultra cold atoms. Here we point out that this lofty goal can be achieved by using a simple scheme of shaking an optical lattice, which is directly applicable in current experiments.  The photon-assistant band hybridization leads to the production of an effective spin-orbit coupling, in which the band index represents the pseudospin. When this spin-orbit coupling has finite strengths along multiple directions, non-trivial topological structures emerge in the Brillouin zone, such as topological defects with a winding number 1 or 2 in a shaken square lattice. The shaken lattice also allows one to study the transition between two band structures with distinct topological properties. 


\end{abstract}

\maketitle

The study on topological matters is one of the most important themes in condense matter physics in the past few years\cite{Kane,  ZhangSC}.  When non-trivial topology exists in the band structures of certain solid materials, a wide range of novel topological matters arise. Whereas the effort of searching for such materials in solids has been continuously growing,  there have been great interests of realizing topological matters using ultra cold atoms\cite{Wang, Dalibard, Essin, Goldman, Liu}.  In such highly controllable atomic systems, it is easy to manipulate the interaction between atoms and external fields so that  topological properties of quantum matters could be engineered using standard experimental techniques. It is hoped that ultra cold atoms will not only provide a perfect simulator of electronic systems, but also opportunities to create new types of topological matters with no counterpart in solids. 

As SOC is a key ingredient in many topological matters, the realization of synthetic  spin-orbit coupling(SOC) using the Raman scheme\cite{Ian,  Martin, Chen, Zhang, ChenY, Engels} opens the door for accessing topological matters in ultra cold atoms. However, a shortcoming of the current scheme is that SOC exists along only one spatial direction.  This has become one of the bottlenecks for an experimental realization of topological matters in ultra cold atoms. 


Both theoretical and experimental interests on shaken optical lattices have been arising recently\cite{Sengstock, Sengstock2, Chin, Windpassinger, Hauke, Koghee, Esmann}. It has been shown that such a scheme allows one to manipulate both the magnitude and the sign of tunneling constants. 
In this Letter, we point out that shaken lattices provide physicists an unprecedented opportunity to explore topological matters. We will show that (I) one could use shaken lattices to create a fully controllable ``SOC" with finite strengths along multiple spatial directions, where band indices play the role of the ``spin" degree of freedom; (II) such an effective SOC allows one to create band structures with non-trivial topological properties using currently available experimental techniques; (III) varying these microscopic parameters, including the frequency, amplitude and phase shift of the shaken lattice,  physicists could study the evolution between two band structures with different topological properties.

As an example,  we show that a two-dimensional shaken square lattice, where  photon-assistant band hybridization creates an effective SOC near the $\Gamma$(${\vec k}=(0,0)$) and $M$($\vec{k}=(\pi,\pi)$) point in the Brillouin zone(BZ), 
\begin{equation}
H=A(k_x^2-k_y^2)\sigma_z+(B k_xk_y+C)\sigma_x+D\sigma_y\label{MO}, 
\end{equation}
where $A, B, C, D$ are momentum-independent constants.  $A\sim (t_p+ t_s)/4$ is determined by the static lattice, $B$ can be regarded as the strength of a momentum-dependent magnetic field in the transverse direction,  $C, D$ correspond to the strengths of a constant magnetic field. Depending on the choices of these parameters, which are well tunable in shaken lattices, the Hamiltonian in Eq.(\ref{MO}) can be classified to two categories. 

{\it Case 1} $B=0$.  Eq.({\ref{MO}) reduces to $H=A(k_x^2-k_y^2)\sigma_z+C\sigma_x+D\sigma_y$,  where spin-momentum locking exists along only one direction,  similar to SOC realized by the Raman scheme in continuum\cite{Ian,  Martin, Chen, Zhang, ChenY, Engels}. This type of SOC does not give rise to interesting topological properties of the band structure. 

{\it Case 2} $B\neq 0$. SOC exists along multiple spatial directions and leads to nontrivial topological properties of band structures. A special case is $C=D=0$, which corresponds to a SOC of $d$-wave nature, as $H=Ak^2 \cos(2\theta_{\bf k})\sigma_z+Bk^2\sin(2\theta_{\bf k})\sigma_x$, where $\theta_{\bf k}=\arg\{k_x+ik_y\}$.  It could be used to produce a topological semimetal\cite{Sun}, and is also relevant in the studies of crystalline topological insulators\cite{Fu}. 
  
The lattice we consider is written as
\begin{equation}
V({\bf r}, t)=V\sum_{i=x,y} \cos^2(k_0r_i+f\cos(\omega t+\varphi_i)/{2})+V'({\bf r})\label{Vt}
\end{equation}
where $r_x=x$, $r_y=y$, $k_0=\pi/d$, $d$ is the lattice spacing. $f$ is the shaking amplitude, $\omega$ is the frequency, and $\varphi_i$ is the phase of the shaking along the $x,y$ directions, as shown in Fig.(1A). An additional lattice $V'({\bf r})=\alpha V\cos(2k_0x)\cos(2k_0y)$ is introduced to make the external potential to be inseparable(See Supplementary Materials), where $\alpha$ is a small number. We set $\varphi_x=0$ and $\varphi_y=\varphi$.  $\varphi=0$ and $\varphi=\pi/2$ correspond to a linear shaking along the diagonal direction and a cyclic mode respectively. 



Eq.(\ref{Vt}) can be regarded as Kramer-Henneberger representation of an irradiated lattice\cite{IL,IL2}. Using standard Floquet-Bloch theorem, the solution of the Schrodinger equation could be written as $\Psi({\bf r},t)=e^{-i\epsilon t}\Phi_{\bf k}({\bf r},t)=e^{-i\epsilon t}e^{i{\bf k}\cdot{\bf r}}u_{\bf k}({\bf r},t)$, where $\epsilon$ is the quasienergy, $u_{{\bf k}}({\bf r},t)$ satisfies $u_{\bf k}({\bf r+R},t+2\pi/\omega)=u_{\bf k}({\bf r},t)$, and ${\bf R}$ is the lattice vector of the static one. Eq.(\ref{Vt}) possesses a certain spatio-temporal(dynamical) symmetry, which allows one to conclude that for the circular and linear shaking, Floquet-Bloch bands have a four- and two-fold symmetry respectively (See Supplementary Materials). 

 
 \begin{figure}[tbp]
\begin{center}
\includegraphics[width=3.1in]{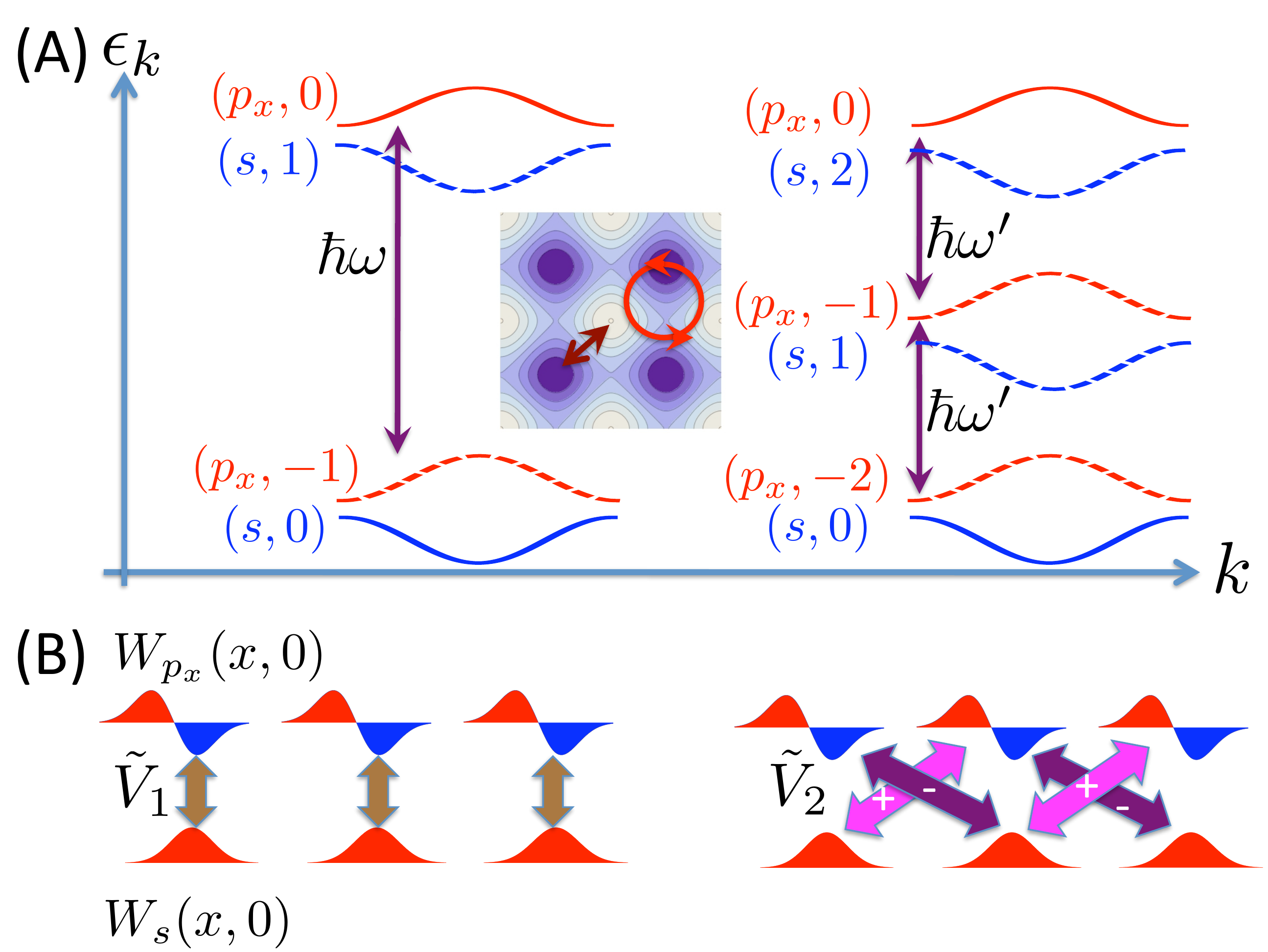}
\end{center}
\caption{(A) Photon-assistant band hybridization. Solid and dashed curves represent bands in the static lattice and side bands produced by shaking. In the left panel, $\omega$ is close to the separation between the $s$ and $p$ bands of the static lattice, while the frequency $\omega'$ for the right pannel is half of $\omega$. Inset is a schematic of the shaken lattice.  The arrow and circle represent the linear and cyclic shaking respectively. (B) Due to parity conservation, $1-$ photon resonance couples orbital with different parities at the same lattice site, whereas $2-$ photon resonance couples orbital with different parities at the nearest neighbor sites.  Red and blue colors represent signs of the wave functions. }
 \end{figure}

Applying the identify for Bessel function $\exp[{x(\zeta-\zeta^{-1})/2}]=\sum_{n=-\infty}^{\infty} J_n(x)\zeta^n$, we obtain $V({\bf r}, t)=V_0({\bf r})+V'({\bf r})+\sum_{n\neq 0}V_n({\bf r})e^{in\omega t}$, 
\begin{equation}
V_n({\bf r})=\bigg\{
\begin{array}{cc}
\frac{i^n}{2} VJ_n(f)&\left( \cos (2k_0 x)+e^{in\varphi} \cos(2k_0y)\right), \\
& n\in even\\
\frac{i^{n+1}}{2} VJ_n(f)& \left(\sin(2k_0 x)+e^{in\varphi}\sin(2k_0y)\right). \\
& n\in odd
\end{array}\label{Vn}
\end{equation}
where $V_0({\bf r})+V'({\bf r})$ is time-independent,  and ${V}_n({\bf r})$ is a dynamically induced lattice potential that excites the system by a multiple-photon energy $n\hbar\omega$, as shown in Fig (1B).  Eq.(\ref{Vn}) shows that $\omega$ controls which bands shall be hybridized at resonance, i.e., which ${V}_n({\bf r})$ is dominant. Moreover, it shows  that the parity of $V_n({\bf r})$ is $(-1)^n$, which gives rise to distinct properties of the band hybridization for even and odd values of $n$. This can be directly seen in the tight-binding picture. As shown in Fig (1B), due to the parity conservation, ${V}_n({\bf r})$ with an even $n$ cannot couple Wannier wave functions with different parities, for instance, $s$ and $p_x$ orbitals,  at the same lattice site. Its leading contribution is to couple them at nearest neighbor sites. Moreover, the coupling between a $s$ orbital at site ${\bf R}_i $with the two $p_x$ orbitals at ${\bf R}_i\pm d\hat{x}$, where $\hat{x}$ is a unit vector along the x direction, differ by a minus sign. This provides a momentum-dependent inter-band coupling for accessing nontrivial topological band structures. In contrast, for ${V}_n({\bf r})$ with an odd $n$, it couples Wannier wave functions with different parities at the same site. Its leading contribution is a momentum-independent inter-band coupling, which could not produce non-trivial topological bands.


 \begin{figure}[tbp]
\begin{center}
\includegraphics[width=3.2in]{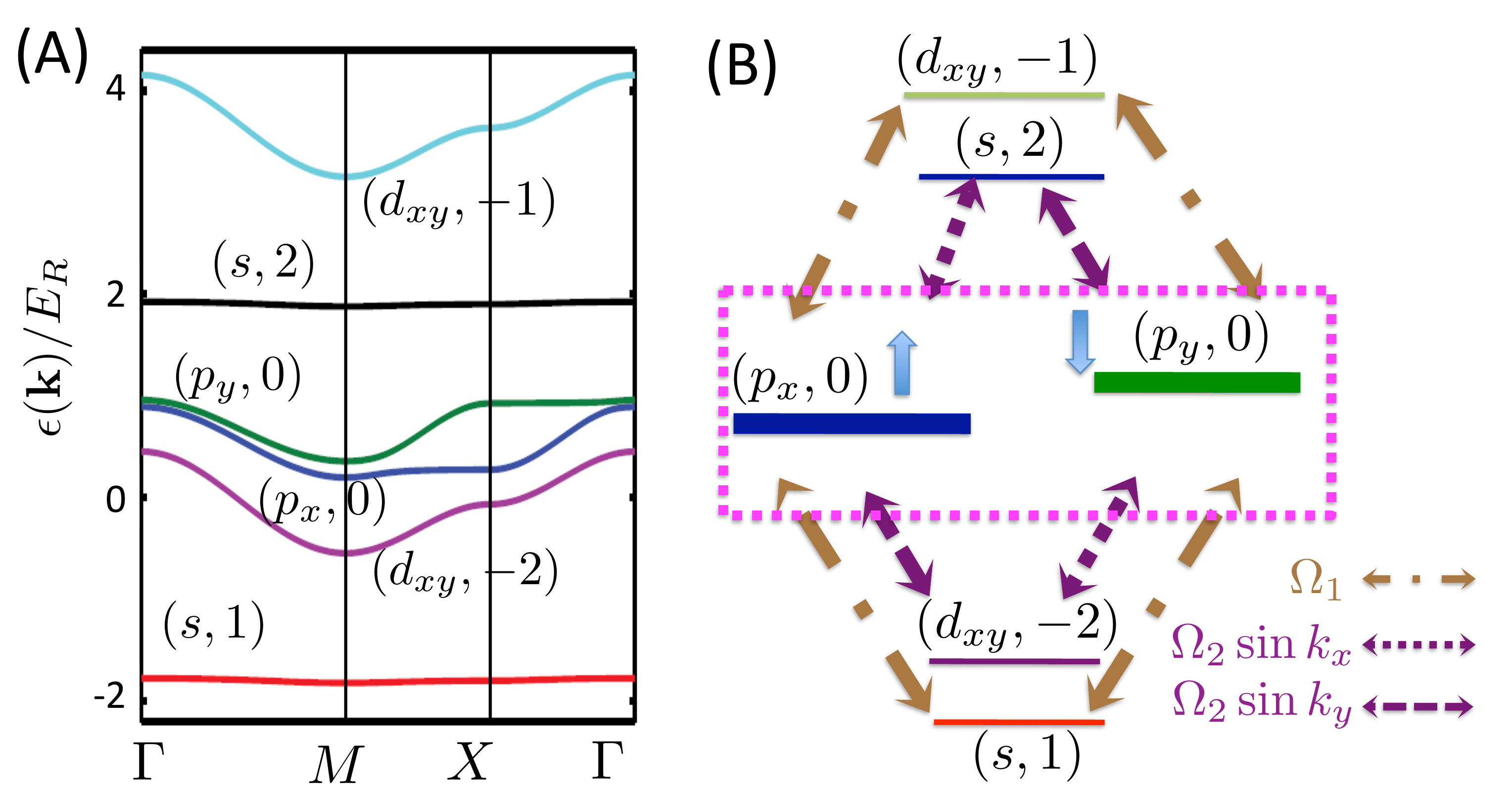}
\end{center}
\caption{(A) A typical band structure, where  $V=16E_R$, $E_R=\hbar^2\pi^2/(2md^2)$, $\alpha=0.2$, $f=0.1d$, $\omega=3.7E_R$.  (B) Schematic of the effective SOC produced by photon-assistent band hybridization.  The $(p_x, 0)$, $(p_y, 0)$ bands (highlighted in the dashed box) form a pseudo--spin-1/2 system. Each of the four bands $(s,1)$, $(s,2)$, $(d_{xy},-1)$ and $(d_{xy},-2)$ serves as an intermediate state as the one in a typical $\Lambda$ transitions. Due to the parity conservation as shown in Fig (1), $(s,2)$ and $(d_{xy,-2})$ band naturally provide a momentum dependent coupling between $(p_x, 0)$ and $(p_y, 0)$ while $(s,1)$ and $(d_{xy},-1)$ only provide a constant coupling. }
 \end{figure}



We expand the Floquet mode as $\Phi_{\bf k}({\bf r},t)=\sum_{m{\bf k},n} c_{m {\bf k}, n} \phi_{m {\bf k}}({\bf r}) e^{in\omega t}$, where $c_{m {\bf k}, n} $ are time-independent constants, and $\phi_{m {\bf k}}({\bf r})$ is the Bloch wave function of the static lattice $V_0({\bf r})$ with band index $m$ and crystal momentum ${\bf k}$. The standard Floquet-matrix representation may be expressed as
\begin{equation}
\begin{split}
\sum_{m', n'}(\mathcal{V}^{m,m'}_{n-n',{\bf k}}+\mathcal{V'}^{m,m'}_{{\bf k}}\delta_{n,n'} +(\epsilon_{m'{\bf k}}^0\\+n'\hbar\omega)\delta_{n,n'}\delta_{m,m'})    c_{m' {\bf k}, n'}   =\epsilon  c_{m {\bf k}, n},\label{FE}
\end{split}
\end{equation}
where $\mathcal{V}^{m,m'}_{n-n',{\bf k}}=\int d{\bf r}\phi_{m{\bf k}}^*({\bf r})V_{n-n'}({\bf r})\phi_{m'{\bf k}}({\bf r})$ and $\mathcal{V'}^{m,m'}_{{\bf k}}=\int d{\bf r}\phi_{m{\bf k}}^*({\bf r})V'({\bf r})\phi_{m'{\bf k}}({\bf r})$. 
The physical meaning of Eq.(\ref{FE}) is apparent. A band of the static lattice could absorb or emit $n$ photons and form a sequence of side bands.  This photon-assistant process make a resonance between certain side bands possible. For convenience, we use the notation $(m, n)$ to represent the dynamically generated $n$th side band of the band $m$ of the static lattice. The coupling between two side bands through $\mathcal{V}^{m,m'}_{n,{\bf k}}$ will be referred as to a $n$-photon process. 

As discussed before, the matrix elements $\mathcal{V}^{m,m'}_{n,{\bf k}}$ is either a constant or ${\bf k}$-dependent,  depending on $n$ and the parity difference of these two bands . For instance, 
\begin{equation}
\begin{split}
&\mathcal{V}^{s,p_x}_{2l,{\bf k}}=i^{2l+1}\Omega_{2l}\sin{k_x}, \mathcal{V}^{s,p_y}_{2l,{\bf k}}=i^{2l+1} e^{2il \varphi}\Omega_{2l}\sin{k_y},\\
&\mathcal{V}^{s,p_x}_{2l+1,{\bf k}}=i^{2l+2}\Omega_{2l+1}, \mathcal{V}^{s,p_y}_{2l+1,{\bf k}}=i^{2l+2}e^{i(2l+1)\varphi}\Omega_{2l+1}\\
\end{split}
\end{equation}
where $l$ is an integer, and $d$ has been absorbed to $k_{i=x,y}$, $\Omega_{2l}=VJ_{2l}(f) \langle W_{s, {\bf R}_i} |\cos(2k_0x)|W_{p_x, {\bf R}_i+d\hat{x}}\rangle$, $\Omega_{2l+1}= \frac{V}{2}J_{2l+1}(f)\langle W_{s, {\bf R}_i} |\sin(2k_0x)|W_{p_x, {\bf R}_i}\rangle$ are constants,  $W_{m, {\bf R}_i}$ is a Wannier function for the band $m$ at the lattice site ${\bf R}_i$. Other matrix elements are provided in the Supplementary Materials. 






We perform a numerical calculation on the Floquet-matrix by including up to the $g$ bands, each of which contains $9$ side bands. 
 Tight-binding model for the dispersions in the static lattice has been used, i.e., $\epsilon^0_{s, \bf k}=-t_s(\cos{k_x}+\cos{k_y})-\Delta$, $\epsilon^0_{p_x, \bf k}=t_p\cos{k_x}-t_s\cos{k_y}$, $\epsilon^0_{p_y, \bf k}=-t_s\cos{k_x}+t_p\cos{k_y}$, $\epsilon^0_{d_{xy}, \bf k}=t_p(\cos{k_x}+\cos{k_y})+\Delta$  
 where  $t_s$, $t_p$ are the tunneling amplitudes.  A typical band structure is shown in Fig (2 A). At $\Gamma$ and $M$ point, there are two nearly degenerate bands, the main contributions to which come from  $(p_x, 0 )$ and $(p_y, 0)$. The hybridization with the other side bands lifts the degeneracy at $\Gamma$ and $M$ point present in an ordinary static square lattice. 
 

From the numerical solutions, we have found that the qualitative physics of the Floquet-matrix is captured by a six-band model. As shown in Fig (2 B), each of the four bands $(s, -2)$, $(s, -1)$, $(d_{xy},-1)$ and $(d_{xy},-2)$, couples to both $(p_x,0)$ and $(p_y,0)$, producing a second order virtual hopping processes between the latter two bands, similar to the standard $\Lambda$ process in atomic physics. Including other bands only leads to quantitative changes of the results. If one treats the nearly degenerate $p_y$ and $p_x$ bands as spin-up and spin-down, an effective SOC Hamiltonian can be formulated, $H={\bf B_k}\cdot \vec{\sigma}$, where ${\sigma}_{x,y,z}$ are the Pauli matrices. To be explicit, we obtain
\begin{equation}
\begin{split}
H&=-(t_p+t_s)(\cos{k_x}-\cos{k_y})\sigma_z/2\\
&+(B_{x,e}\sin{k_x}\sin{k_y}+B_{x,o})\sigma_x\\
&+(B_{y,e}\sin{k_x}\sin{k_y}+B_{y,o})\sigma_y,\label{Heff}
\end{split}
\end{equation}
where the subscript $e$ and $o$ represent the effective magnetic field induced by processes of even and odd number of photons.  For the $\sigma_z$ term, the main contribution comes from the energy difference between $(p_x, 0)$ and $(p_y, 0)$ bands, and a small correction $B_z'$ from hybridization with other bands does not affect the results(See Supplementary Materials). For the transverse fields, we define $\bar{B}_{i=e,o}=B_{x,i}-i B_{y,i}$ for convenience, where
\begin{equation}
\begin{split}
\bar{B}_{e}=\Omega_0+\sum_{n\in even}\frac{-e^{-in\varphi}\Omega^2_{n}}{\epsilon_{s, {\bf k}}+n\hbar\omega-(\epsilon_{p_x, {\bf k}}+\epsilon_{p_y, {\bf k}})/2},\\
-\frac{e^{i n\varphi}\Omega^2_{n}}{\epsilon_{d_{xy}, {\bf k}}+n\hbar\omega-(\epsilon_{p_x, {\bf k}}+\epsilon_{p_y, {\bf k}})/2}.\\
\end{split}\label{Bxy}
\end{equation}
where $\Omega_0=\langle W_{p_x,{\bf R}_i+d\hat{y}}|V'({\bf r})|W_{p_y,{\bf R}_i+d\hat{x}}\rangle$, $\epsilon_{m, {\bf k}}=\epsilon^0_{m, {\bf k}}+\langle \phi_{m{\bf k}}({\bf r}) |V'({\bf r})|\phi_{m{\bf k}}({\bf r}) \rangle $ has taken into account the shift of each bands due to $V'({\bf r})$. As for  $B_{x,o}$ and $B_{y,o}$, the expressions are identical, with the summation over odd integers. In the leading order, $\bar{B}_{e,o}$ are momentum independent constants, as $\epsilon_{m, {\bf k}}$ in the denominators of Eq.(\ref{Bxy}) may be replaced by their values at the $\Gamma$ and $M$ point in the numerator. 

To simplify the expressions, we apply a spin rotation about the $z$ axis, $ e^{i\theta \sigma_z/2} \sigma_x e^{-i\theta \sigma_z/2}=\cos\theta \sigma_x+\sin\theta\sigma_y $, $ e^{i\theta \sigma_z/2} \sigma_y e^{-i\theta \sigma_z/2}=-\sin\theta\sigma_x+\cos\theta \sigma_y$, where $\tan(\theta)=-B_{y,e}/B_{x,e}$, so that $B_{x, e}\sigma_x+B_{y, e}\sigma_y \rightarrow B_{e}\sigma_x$, where $B_e=\sqrt{B^{2}_{x,e}+B^2_{y,e}}$. Near the $\Gamma$ point, $\cos{k_x}-\cos{k_y}\sim (k_x^2-k_y^2)$, the effective magnetic field ${\bf B_k}$ in Eq.(\ref{Heff})becomes 
\begin{equation}
\begin{split}
{\bf B_k}=\Big(B_{e}k_xk_y+\tilde{B}_{x,o}, \tilde{B}_{y,o}, \frac{t_p+t_s}{4}(k_x^2-k_y^2)\Big)
\end{split}\label{Bf}
\end{equation}
where $\tilde{B}_{x,o}=B_{x,o}\cos\theta-B_{y,o}\sin\theta$ and $\tilde{B}_{y,o}=B_{x,o}\sin\theta+B_{y,o}\cos\theta$. This leads  to the expression for the Hamiltonian in Eq.($\ref{MO}$).  The formalism of the Hamiltonian near the $M$ point is the same, with quantitatively different values of the three components of ${\bf B_k}$.  Depending on the choice of $\omega$ and $f$, both {\it Case 1} and {\it Case 2} of Eq.(\ref{MO}) can be realized.  




{\it 1-photon process}  
If one tunes the $(s,1)$ band to be closest to the $(p_x,0), (p_y,0)$ bands, the 1-photon process is dominant, which leads to $\tilde{B}_{x,o}\gg B_{e}$.  Eq.(\ref{Heff}) becomes 
\begin{equation}
H=\pm\frac{t_s+t_p}{4}(k_x^2-k_y^2)\sigma_z+\tilde{B}_{x,o}\sigma_x+\tilde{B}_{y,o}\sigma_y, 
\end{equation} 
where $\pm$ corresponds to the $\Gamma$ and $M$ points respectively. {\it Case 1} of Eq. (\ref{MO}) is then achieved.

{\it 2-photon process} 
Choosing a proper frequency so that $(s, 2)$ is the closest one to $(p_x,0)$ and $(p_y,0)$ bands,  2-photon process is dominate.  It is worth pointing out that $\Omega_2$ can be further amplified by significantly enlarging the overlap integral for the Wannier wave functions in the nearest neighbor sites in double-well lattices\cite{Zhou}.  Eq.(\ref{Heff}) then becomes 
\begin{equation}
H=\pm\frac{t_s+t_p}{4}(k_x^2-k_y^2)\sigma_z+B_{e}k_xk_y\sigma_x 
\end{equation} 
Topological defects then emerge at  the $\Gamma$ and $M$ points where the effective magnetic field ${\bf B_k}$ vanishes. For a closed loop in the momentum space around one of these two points , a winding number of $\pm 2$ of ${\bf B_k}$ is evident, as ${\bf B_k}\sim (\sin(2\theta_{\bf k}),0, \cos(2\theta_{\bf k}))$. 

In general, both 1- and 2-photon processes contribute to the effective Hamiltonian. Eq. (\ref{Bf}) allows one to investigate how the two band structures with distinct topological properties may evolve from one to the other when  $\omega$ continuously changes.  From the numerical solution of the Floquet-matrix, we find that if the $(p_x, 0)$ and $(p_y,0)$ bands are not degenerate with other side bands at the $\Gamma$ and $M$ point, a spin-$1/2$ description is sufficient for describing the eigenstates near these two points, as they are dominated by $(p_x, 0)$ and $(p_y,0)$ bands.  The spin eigen state is written as $(\cos (\alpha_{\bf k}/2),  e^{i\beta_{\bf k}}\sin(\alpha_{\bf k}/2))$, from which an effective magnetic field ${\bf B_k}$ is constructed, as $\alpha_{\bf k}$ and $\beta_{\bf k}$ correspond to the direction of the unit vector ${\bf B_k}/|{\bf B_k}|$ on the Bloch sphere and the energy splitting gives rise to the strength of ${\bf B_k}$. 

\begin{figure}[tbp]
\begin{center}
\includegraphics[width=3.3in]{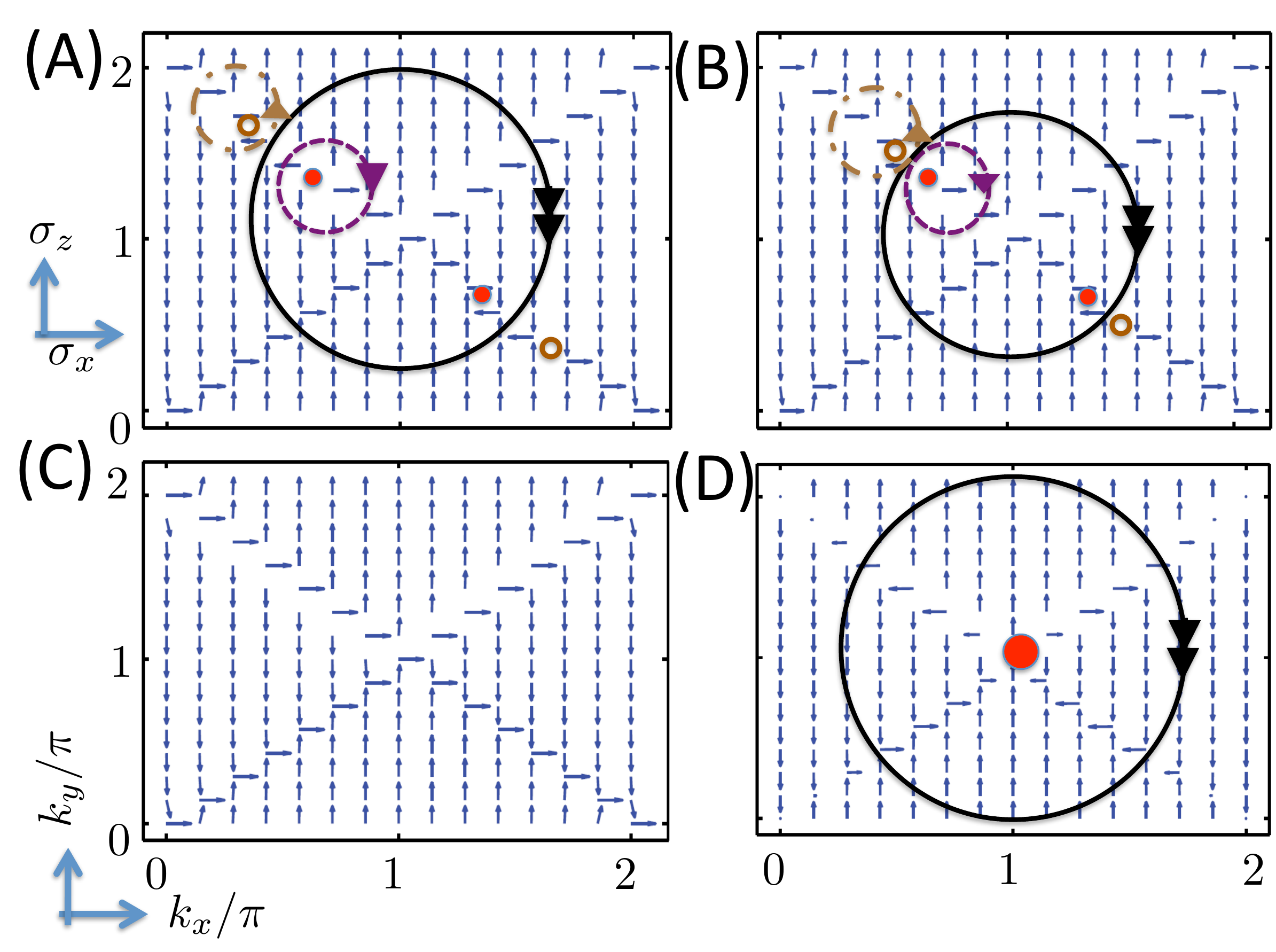}
\end{center}
\caption{Topological defects in the band structure. Small arrows represent the strength and direction of the $x$ and $z$ components of the ${\bf B_k}$ field in Eq.(\ref{Heff}). Such a momentum-dependent magnetic field determines the orientation of the pseudo-spin-1/2 composed by $(p_x,0)$ and $(p_y, 0)$. Topological defects emerge at places where ${\bf B_k}$ vanishes. Small filled and empty dots represent defects with winding number $1$ and $-1$ respectively. The number and directions of big arrows on closed loops represent the winding number of the ${\bf B_k}$ field along these loops in the Brillouin zone. (A-B) For linear shaking with  $\varphi=0$, a finite constant component $\tilde{B}_{x,o}$ in Eq.(\ref{Bf}) leads to the splitting of the defects of winding number 2 to two defects with winding number 1. (C) Topological defects with opposite winding numbers have annihilated each other when the shaken frequency excesses  a certain critical value. (D)  For circular shaking with $\varphi=\pi/2$,  $\tilde{B}_{x,o}=0$, the spin is tilted towards the normal direction of  the $k_x-k_y$ plane due to a finite $\tilde{B}_{y,o}$ (not shown). The projection of the spin on the plane has a winding number of 2. With increasing $\tilde{B}_{y,o}$, all spins eventually become perpendicular to the plane. For all figures, $V=20E_R$, $\alpha=0.05$, $f=0.03d$. $\omega/E_R=3.9, 4.0,4.2,4.2$ for (A-D). }
 \end{figure}

Fig. (3) shows a few typical topological structures of ${\bf B_k}$. When $\varphi=0$, ${\bf B_k}$ has only the $x$ and $z$ components.  When $(s, 2)$ is the closest side band to $(p_x, 0)$ and $(p_y,0)$ bands,  topological defects are present. Interestingly, we find that, due to a finite $\tilde{B}_{x, o}$ induced by the $(s, 1)$ band, the topological defects of winding number 2 at the $\Gamma$ point splits to two ones with winding number 1 as shown in Fig (3 A). The same phenomenon occurs at the $M$ point. This can be seen from the fact that ${\bf B}$ now vanishes at $(\pm k^*, \mp k^*)$, where $k^*=\sqrt{\tilde{B}_{x,o}/B_e}$ and $\tilde{B}_{x,o}/B_e>0$ in our case. Near these two points, ${\bf B_k} \sim(\tilde{k}_x+\tilde{k}_y, 0, \tilde{k}_x-\tilde{k}_y)$ that corresponds to a winding number 1, where $\tilde{k}_{i=x,y}=k_{i}\pm k^*$. We have also verified that if $\Omega_1=0$, the splitting is absent and only defects of winding number 2 show up at the $\Gamma$ and $M$ points.  In general cases with a finite $\tilde{B}_{x, o}$,  the winding number of the ${\bf B}$ field on a closed loop in the BZ depends on how many defects it encloses, as shown in Fig. (3 A). If one makes $(s, 1)$ to be more close to the $p$ bands with changing $\omega$, $\tilde{B}_{x, o}$ and $k^*$ increases, and the defect with winding number 1 split from the $\Gamma$ point gradually approaches the defect  of winding number $-1$ from the $M$ point, as shown Fig. (3 B, C), and the topological structures eventually disappears. This establishes the evolution between two band structures with distinct topological properties. As the spin corresponds to the band index, the topological structure here and its evolution can be visualized in experiments using a variety of schemes\cite{Esslinger, Zhao, Seeing}.




 


It is worth pointing out that $\tilde{B}_{x,o}$ relies on the phase shift $\varphi$. For the cyclic shaking,  $\varphi=\pi/2$, $\tilde{B}_{x,o}=0$.  Changing the value of $\omega$ only leads to a tilting of the spin along the $y$ direction, and the winding number of the spin on the $\sigma_x-\sigma_z$ plane is not affected, as shown in Fig. (3 D). This also indicates that for the cyclic shaking, there is a four-fold symmetry in the Brillouin zone. We have verified that figure 3 (A-D) are consistent with general understandings of the symmetry of  the Floquet-Bloch bands for linear and circularly shaking lattice as discussed before.

Whereas we have been focusing on two dimensions  in this Letter, all discussions can be directly applied to one dimension, where a shaken lattice produces nearly flat bands with a Zak phase  $\pm \pi$(See Supplementary Materials). The general principle of producing a multi-dimensional SOC using dynamically generated band hybridization could be straightforwardly generalized to other lattice geometries. In practice, to minimize heating effects, one should choose a small shaken amplitude, and a shaking frequency that is off-resonance with characteristic energy scales in a single Floquet-Bloch band such as $t_s$ and $t_p$. In C. Chin's recent experiment\cite{Chin}, both these requirements have been fulfilled, and  atoms can be prepared in a desired Floquet-Bloch band with long lifetime up to 1s.  Whereas a one-photon resonance has been used in current experiments, one could straightforwardly generalize such a shaken scheme to two-photon resonance. It is very promising that the interplay between tunable lattice geometry and the well controllable shaking scheme will lead to fruitful results on shaping the topology of band structures in optical lattices in the near future. 

{\it Acknowledgement } This work is supported by NSFC-RGC( NCUHK453/13). 

{\it Note} Near the completion of this manuscript, two preprints (arXiv:1402.3295, arXiv:1402.4034) on topological band structures in shaken optical lattices have just appeared.

\vspace{0.2in}

{\bf \Large Supplementary Material}

{\bf Spatio-temporal (dynamical) symmetry of shaken lattices}

Consider a static lattice with a four-fold symmetry, i.e., the one discussed in the main text. If one circularly  shakes it, i.e., $\varphi=\pi/2$, the Floquet-Bloch Hamiltonian $e^{-i{\bf k}\cdot{\bf r}}(\hat{P}^2/2m+V({\bf r}, t))e^{i{\bf k}\cdot{\bf r}}$ remains unchanged under a simultaneous $\pi/2$ rotation in the real and momentum space and a time-translation, \begin{equation} x\rightarrow y, y\rightarrow -x, k_x\rightarrow k_y, k_y\rightarrow -k_x, t\rightarrow t-\pi/(2\omega). \end{equation} Floquet-Bloch bands are four-fold symmetry. In contrast, for linear shaking with $\varphi=0$, the Floquet-Bloch Hamiltonian is no longer invariant under the above transformation. Instead, it is unchanged only if ${\bf r}\rightarrow -{\bf r}$, ${\bf k}\rightarrow -{\bf k}$ and $t\rightarrow t-\pi/\omega$, which shows that Floquet-Bloch bands have two-fold symmetry under linear shaking. Different symmetries of Floquet-Bloch bands indicate different topological structures under circularly or linearly shaking, as shown in Fig. (3) of the main text.  

Whereas the above discussions apply to any lattices with a finite value of $\alpha$, the case that $\alpha=0 $ and $V'({\bf r})=0$ needs a special consideration. Under this situation,  the lattice potential becomes separable, i.e., $V({\bf r}, t)=V(x, t)+V(y,t)$, the Floquet equation $(H(x,y,t)-i\partial_t)\Phi_{\bf k}(x,y,t)=\epsilon \Phi_{\bf k}(x,y,t)$ can be decoupled to two independent equations
\begin{eqnarray}
\begin{split}
\left(-\frac{\hbar}{2m}\partial_x^2+V\cos^2(k_0x+f\cos (\omega t))-i\partial_t\right)\Phi_{k_x}(x,t)\\=\epsilon_{k_x} \Phi_{k_x}(x,t)\\
\left(-\frac{\hbar}{2m}\partial_y^2+V\cos^2(k_0x+f\cos (\omega t+\varphi)-i\partial_t)\right)\Phi_{k_y}(y,t)\\=\epsilon_{k_y} \Phi_{k_y}(y,t), 
\end{split}
\end{eqnarray}
where 
\begin{eqnarray}
&\Phi_{\bf k}(x,y,t)=\Phi_{k_x}(x,t)\Phi_{k_y}(y,t), \,\,\,\,\,\,\,\,\, \epsilon_{\bf k}=\epsilon_{k_x}+\epsilon_{k_y}. 
\end{eqnarray}

\begin{figure}[tbp]
\begin{center}
\includegraphics[width=3.3in]{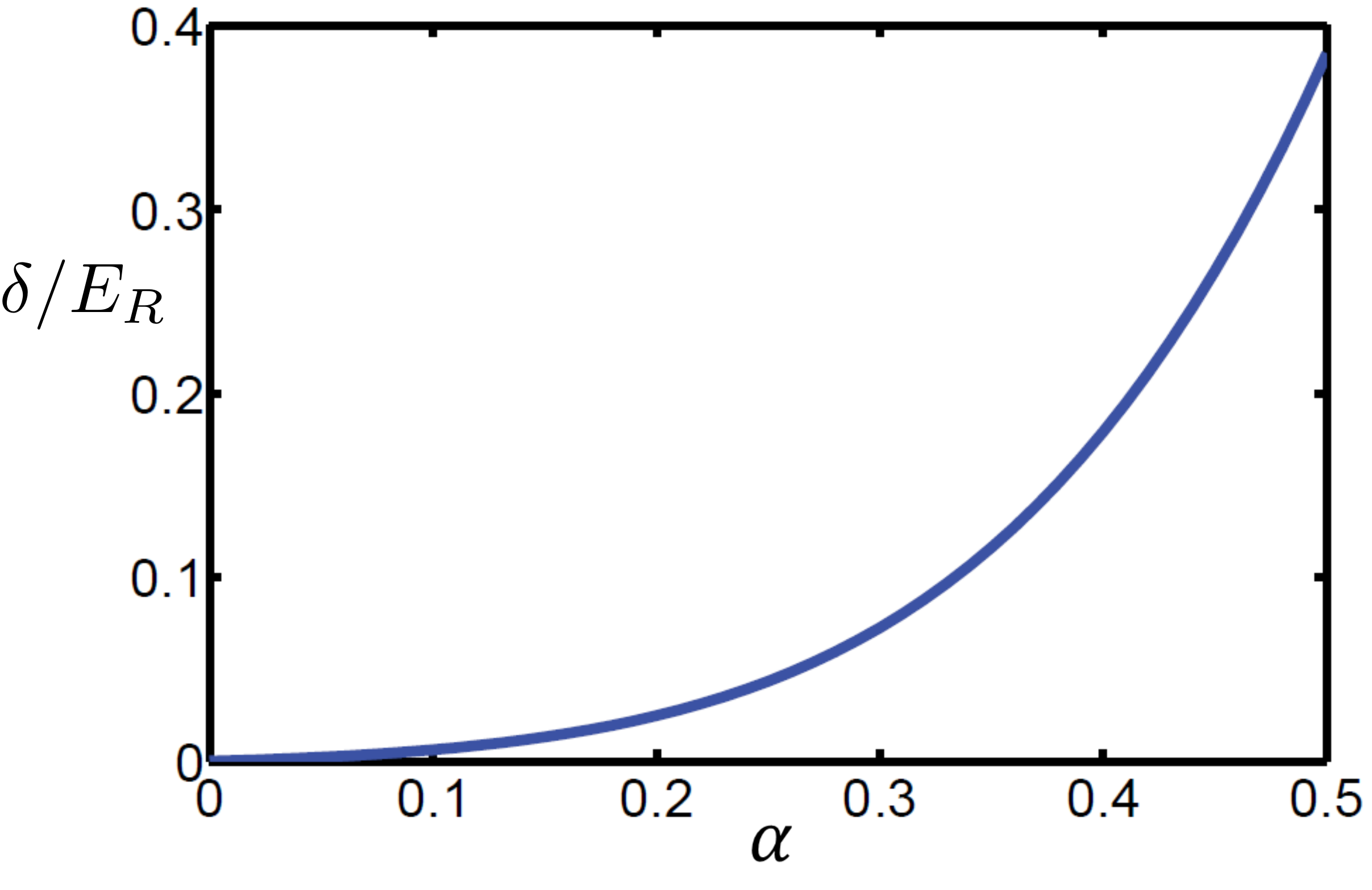}
\end{center}
\caption{The energy difference $\delta$ between the $(p_x,0)$ and the $(p_y,0)$ bands, where $V=16E_R$. As a demonstration, $\delta$ is evaluated at $(\pi/2, \pi/2)$. When $\alpha=0$, the lattice potential becomes separable, $\delta$ vanishes at any points along the $(\pi, \pm \pi)$ directions. For any infinitesimal $\alpha$, $\delta$ becomes finite.  }
 \end{figure}

\begin{figure}[tbp]
\begin{center}
\includegraphics[width=3.3in]{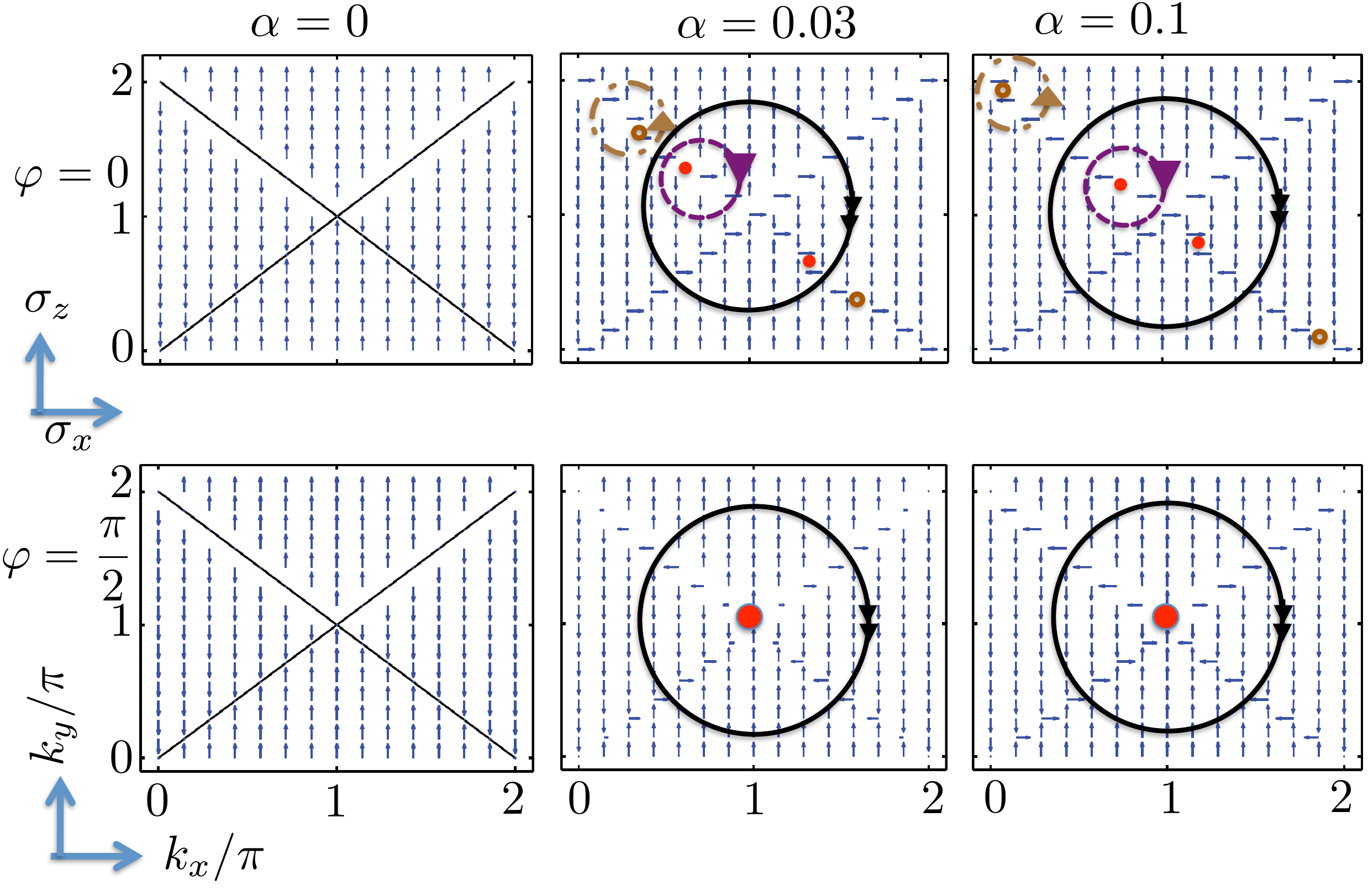}
\end{center}
\caption{Topological band structures as a function of $\alpha$. When $\alpha=0$(left column),  the band structure is independent on $\varphi$. The two bands, $(p_x, 0)$ and $(p_y,0)$ become degenerate along the $(\pi, \pm \pi)$ directions, as indicated by the black solid lines. For any finite values of $\alpha$(middle and right column), $\delta$ becomes finite away from the location of defects. For all figures, $V=20E_R$, $f=0.01d$ and $\omega=8.4E_R$. }
 \end{figure}

It is clear that the phase difference $\varphi$ can be absorbed by applying a time-translation $t\rightarrow t-\varphi/\omega$ to  Eq.(2) alone. The  Floquet-Bloch spectrum is therefore independent on $\varphi$.  In particular, due to Eq.(4), $p_x$ and $p_y$ bands become degenerate if $k_x=\pm k_y$, i.e., along the $(\pi, \pi)$ and $(\pi, -\pi)$ direction in the BZ, as shown in Fig(1) of this supplementary material. 

For any infinitesimal $\alpha\neq 0$,  the Floquet equation is no longer separable, and a finite gap emerges along the $(\pi, \pi)$ and $(\pi, -\pi)$ directions in the BZ. As shown in Fig (2) of this supplementary material, depending on the choice of the phase shift $\varphi$, Floquet-Bloch bands exhibit different symmetries and topological structures. Such a dynamical symmetry was not studied in shaken lattice before.


\vspace{0.2in}

{\bf Matrix elements $\mathcal{V}^{i,j}_{n,{\bf k}}$ }

Take $s$ and $p_x$ band as an example, we expand the Bloch wave functions in the basis of Wannier functions $W_m({\bf r})$, and rewrite $\mathcal{V}^{s,p_x}_{n,{\bf k}}$ as 
\begin{equation}
\mathcal{V}^{s,p_x}_{n,{\bf k}}=\sum_{{\bf R}_i{\bf R}_j}\int d{\bf r}W_s({\bf r-R}_i)W_{p_x}({\bf r-R}_j)V_n({\bf r})e^{i{\bf k\cdot}({\bf R}_j-{\bf R}_i)}.\label{IN}
\end{equation}
As the static state $V_0({\bf r})$ is separable along the $x$ and $y$ directions, the two Wannier wave function can be written as $W_s({\bf r})=w_0(x)w_0(y)$ and $W_{p_x}({\bf r})=w_1(x)w_0(y)$, where $w_0(x)$ and $w_1(x)$ are the lowest two Wannier functions for a one dimensional lattice $V_0(x,0)$ or $V_0(0,x)$ respectively. Apparently, $W_s({\bf r})=W_s(-{\bf r}) $ and $W_{p_x}(x, y)=-W_{p_x}(-x, y)=W_{p_x}(x, -y)$. 
 
If $n$ is odd, one sees that the integral in Eq.(\ref{IN}) is finite when taking $i=j$, due to the fact that $V_{2l+1}({\bf r})=-V_{2l+1}(-{\bf r})$. This means that $V_{2l+1}({\bf r})$ is able to couple the Wannier orbital $W_s({\bf r-R}_i)$ and $W_p({\bf r-R}_i)$ at the same lattice site ${\bf R}_i$. Meanwhile, the integral in Eq.(\ref{IN})  is much smaller if $i\neq j$ because of the the small overlap of the Wannier wave functions at different lattice site. Therefore, $\mathcal{V}^{s,p_x}_{2l+1,{\bf k}}$ becomes a constant in the leading order, 
\begin{equation}
\mathcal{V}^{s,p_x}_{2l+1,{\bf k}}=i^{2l+2}\Omega_{2l+1},\,\,\,\,\,\,\, \mathcal{V}^{s,p_y}_{2l+1,{\bf k}}=i^{2l+2}e^{i(2l+1)\varphi}\Omega_{2l+1}
\end{equation}
where $\Omega_{2l+1}= \frac{V}{2}J_{2l+1}(f)\langle W_{s, {\bf R}_i} |\sin(2k_0x)|W_{p_x, {\bf R}_i}\rangle$. 

If $n$ is even, the situation is very different. It is clear that $V_{2l}({\bf r})$ is not able to couple the two Wannier orbital $W_s({\bf r})$ and $W_p({\bf r})$ at the same lattice site. The leading contribution to $\mathcal{V}^{s,p_x}_{2l,{\bf k}}$ therefore must come from the nearest neighbor ones.  Through a simple calculation, one sees that 
\begin{equation}
\mathcal{V}^{s,p_x}_{2l,{\bf k}}=i^{2l+1}\Omega_{2l}\sin(k_xd), \,\,\,\,\,\,\, \mathcal{V}^{s,p_y}_{2l,{\bf k}}=i^{2l+1} e^{2il \varphi}\Omega_{2l}\sin(k_yd), 
\end{equation}
where $\Omega_{2l}=VJ_{2l}(f) \langle W_{s, {\bf R}_i} |\cos(2k_0x)|W_{p_x, {\bf R}_i+d\hat{x}}\rangle$, $d$ is the lattice spacing and $\hat{\bf x}$ is the unit vector along the $x$ axis. \\

Similarly, we have
\begin{eqnarray}
\mathcal{V}^{p_x,d_{xy}}_{2l,{\bf k}}&=&i^{2l+1}e^{2il \varphi}\Omega_{2l}\sin{k_yd}, \mathcal{V}^{p_y,d_{xy}}_{2l,{\bf k}}=i^{2l+1} \Omega_{2l}\sin{k_xd}\nonumber\\ \mathcal{V}^{p_x,d_{xy}}_{2l+1,{\bf k}}&=&i^{2l+2}e^{i(2l+1)\varphi}\Omega_{2l+1}, \mathcal{V}^{p_y,d_{xy}}_{2l+1,{\bf k}}=i^{2l+2}\Omega_{2l+1}
\end{eqnarray}
\begin{eqnarray}
\mathcal{V}^{d_x,p_x}_{2l+1,{\bf k}}&=&i^{2l+2}\Omega'_{2l+1},\,\,\,\,\,\,\, \mathcal{V}^{d_x,p_y}_{2l+1,{\bf k}}=0\nonumber\\ \mathcal{V}^{d_x,p_x}_{2l,{\bf k}}&=&i^{2l+1}\Omega'_{2l}\sin(k_xd), \,\,\,\,\,\,\, \mathcal{V}^{d_x,p_y}_{2l,{\bf k}}=0\nonumber\\ \mathcal{V}^{d_y,p_x}_{2l+1,{\bf k}}&=&0,\,\,\,\,\,\,\, \mathcal{V}^{d_y,p_y}_{2l+1,{\bf k}}=i^{2l+2}e^{i(2l+1)\varphi} \Omega'_{2l+1}\nonumber\\ \mathcal{V}^{d_y,p_x}_{2l,{\bf k}}&=&0, \,\,\,\,\,\,\, \mathcal{V}^{d_y,p_y}_{2l,{\bf k}}=i^{2l+1} e^{2il \varphi}\Omega'_{2l}\sin(k_yd)
\end{eqnarray}
where $\Omega'_{2l+1}=\frac{V}{2}J_{2l+1}(f)\langle W_{d_x, {\bf R}_i} |\sin(2k_0x)|W_{p_x, {\bf R}_i}\rangle$ and $\Omega'_{2l}=VJ_{2l}(f) \langle W_{d_x, {\bf R}_i} |\cos(2k_0x)|W_{p_x, {\bf R}_i+d\hat{x}}\rangle$.

\begin{eqnarray}
&\mathcal{V}^{s,d_{xy}}_{n,{\bf k}}=0\\ \nonumber
&\mathcal{V}^{s,d_x}_{2l+1,{\bf k}}=i^{2l+2}\Omega_{d,2l+1}\sin(k_xd),\\ \nonumber
&\mathcal{V}^{s,d_y}_{2l+1,{\bf k}}=i^{2l+2}e^{i(2l+1)\varphi}\Omega_{d,2l+1}\sin(k_yd)\nonumber\\ \nonumber
&\mathcal{V}^{s,d_x}_{2l,{\bf k}}=i^{2l+1}\Omega_{d,2l}, \\ \nonumber
&\mathcal{V}^{s,d_y}_{2l,{\bf k}}=i^{2l+1} e^{2il \varphi}\Omega_{d,2l} 
\end{eqnarray}
where $\Omega_{d,2l+1}= VJ_{2l+1}(f)\langle W_{s, {\bf R}_i} |\sin(2k_0x)|W_{d_x, {\bf R}_i+d\hat{x}}\rangle$ and $\Omega_{d,2l}=\frac{V}{2}J_{2l}(f) \langle W_{s, {\bf R}_i} |\cos(2k_0x)|W_{d_x, {\bf R}_i}\rangle$

\vspace{0.2in}

{\bf Matrix elements $\mathcal{V'}^{i,j}_{{\bf k}}$ }

We can define 
\begin{eqnarray}
&\Omega_s=\langle W_{s,{\bf R}_i}|\cos(2k_0x)|W_{s,{\bf_R}_i}\rangle, \\ \nonumber 
&\Omega'_s=\langle W_{s,{\bf R}_i}|\cos(2k_0x)|W_{s,{\bf_R}+d\hat{x}}\rangle, \\ \nonumber
&\Omega_p=\langle W_{p_x,{\bf R}_i}|\cos(2k_0x)|W_{p_x,{\bf_R}_i}\rangle, \\ \nonumber
&\Omega'_p=\langle W_{p_x,{\bf R}_i}|\cos(2k_0x)|W_{p_x,{\bf_R}+d\hat{x}}\rangle, \\ \nonumber
&\Omega_{sp}=\langle W_{s,{\bf R}_i}|\cos(2k_0x)|W_{p_x,{\bf_R}+d\hat{x}}\rangle.
\end{eqnarray}
Then
\begin{eqnarray}
&\mathcal{V'}^{s,s}_{\bf k}=\alpha V \Omega^2_s,\,\,\,\,\,\,\mathcal{V'}^{p_x,p_x}_{\bf k}=\alpha V \Omega_s\Omega_p, \\ \nonumber
&\mathcal{V'}^{p_y,p_y}_{\bf k}=\alpha V \Omega_s\Omega_p,\,\,\,\,\,\,\mathcal{V'}^{d_{xy},d_{xy}}_{\bf k}=\alpha V \Omega^2_p, \\ \nonumber 
&\mathcal{V'}^{s,p_x}_{\bf k}=i\alpha V \Omega_{sp}\Omega_s \sin k_xd, \\ \nonumber
&\mathcal{V'}^{s,p_y}_{\bf k}=i\alpha V \Omega_{sp}\Omega_s \sin k_yd, \\ \nonumber
&\mathcal{V'}^{p_x,d_{xy}}_{\bf k}=i\alpha V \Omega_{sp}\Omega_p \sin k_yd, \\ \nonumber
&\mathcal{V'}^{p_y,d_{xy}}_{\bf k}=i\alpha V \Omega_{sp}\Omega_p \sin k_xd, \\ \nonumber
&\mathcal{V'}^{p_x,p_y}_{\bf k}=\alpha V \Omega^2_{sp}\sin k_xd \sin k_yd, \\ \nonumber
&\mathcal{V'}^{s,d_{xy}}_{\bf k}=-\alpha V \Omega^2_{sp}\sin k_xd \sin k_yd.
\end{eqnarray}

\vspace{0.2in}

{\bf Correction to $B_z$}
\begin{widetext}
\begin{equation}
\begin{split}
\Delta E'=\sum_{n\in even}\left(\frac{\Omega^2_{n}\sin^2(k_y d)}{\epsilon_{s, {\bf k},n}-\epsilon_{p_y, {\bf k}}}-\frac{\Omega^2_{n}\sin^2(k_x d)}{\epsilon_{s, {\bf k},n}-\epsilon_{p_x, {\bf k}}}\right)+\sum_{n\in odd}\left(\frac{\Omega^2_{n}}{\epsilon_{s, {\bf k},n}-\epsilon_{p_y, {\bf k}}}-\frac{\Omega^2_{n}}{\epsilon_{s, {\bf k},n}-\epsilon_{p_x, {\bf k}}}\right)
\nonumber\\+\sum_{n\in even}\left(\frac{\Omega^2_{n}\sin^2(k_x d)}{\epsilon_{d_{xy}, {\bf k},n}-\epsilon_{p_y, {\bf k}}}-\frac{\Omega^2_{n}\sin^2(k_y d)}{\epsilon_{d_{xy}, {\bf k},n}-\epsilon_{p_x, {\bf k}}}\right)+\sum_{n\in odd}\left(\frac{\Omega^2_{n}}{\epsilon_{d_{xy}, {\bf k},n}-\epsilon_{p_y, {\bf k}}}-\frac{\Omega^2_{n}}{\epsilon_{d_{xy}, {\bf k},n}-\epsilon_{p_x, {\bf k}}}\right)\nonumber\\
+\left(\frac{\Omega^2_{0s}\sin^2(k_y d)}{\epsilon_{s, {\bf k}}-\epsilon_{p_y, {\bf k}}}-\frac{\Omega^2_{0s}\sin^2(k_x d)}{\epsilon_{s, {\bf k}}-\epsilon_{p_x, {\bf k}}}\right)
+\left(\frac{\Omega^2_{0p}\sin^2(k_x d)}{\epsilon_{d_{xy}, {\bf k}}-\epsilon_{p_y, {\bf k}}}-\frac{\Omega^2_{0p}\sin^2(k_y d)}{\epsilon_{d_{xy}, {\bf k}}-\epsilon_{p_x, {\bf k}}}\right)
\end{split}
\end{equation}
\end{widetext}

where $n$ in the subscript of  $\epsilon_{s, {\bf k},n}$ and $\epsilon_{p_{xy}, {\bf k},n}$ is the side band index, $\Omega_{0s}=\alpha V\Omega_{sp}\Omega_s$ and $\Omega_{0p}=\alpha V\Omega_{sp}\Omega_p$.  As both these terms $\sim k_x^2-k_y^2$ near the $\Gamma$ and $M$ points, they contribute a correction to the expression of $B_z$. Their coefficients in the parentheses  are much smaller than $t_p+t_s$ in the small $\Omega_n$ limit and can be ignored. 

\vspace{0.2in}
{\bf One-dimensional shaken lattices}

\begin{figure}[bp]
\begin{center}
\includegraphics[width=3.3in]{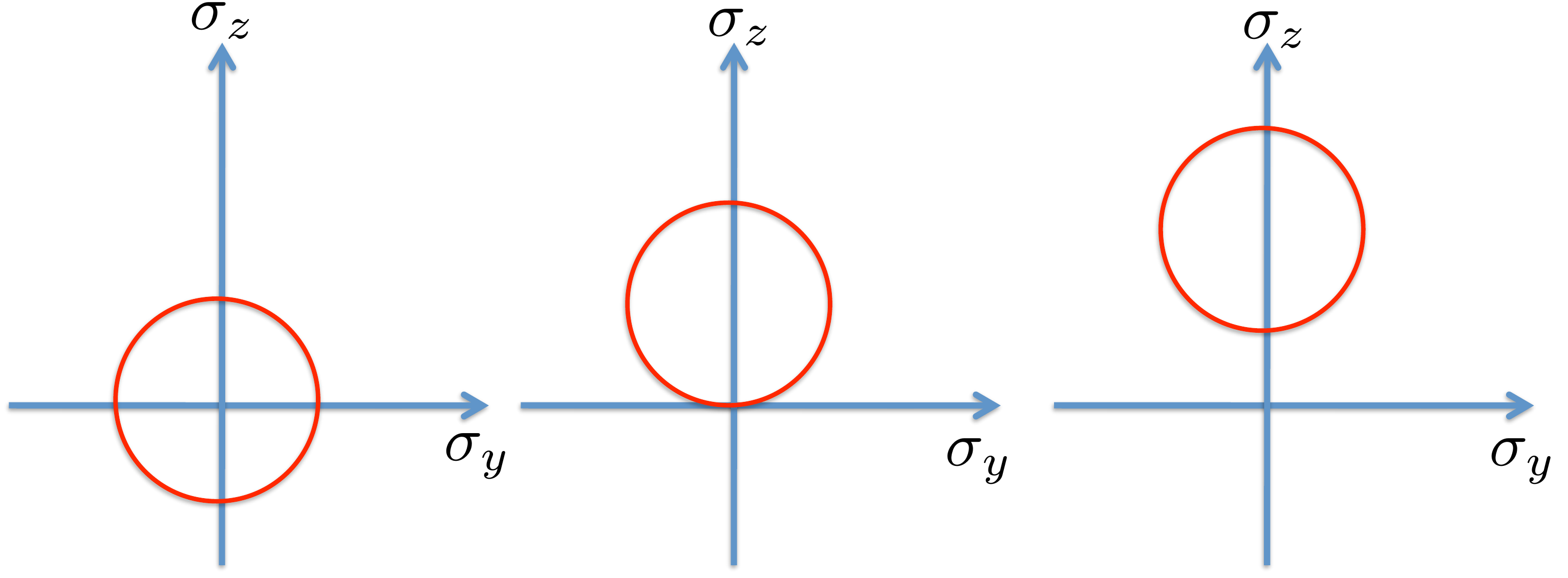}
\end{center}
\caption{Contours of pseudo-spin when the momentum $k_x$ changes from $-\pi/d$ to $\pi/d$. From left to right, the detuning is $0$, $=\tilde{\Delta}_c$ and $>\tilde{\Delta}_c$ respectively.  }
 \end{figure}

All discussions in the main text can be directly generalized to one dimension.  For the 1-photon process that hybridizes $(s,1)$ and $(p, 0)$, the Hamiltonian can be written as $H=(t_s+t_p)\cos(2k_x)\sigma_z/2-\Omega_1\sigma_x$. This is what has been realized in C. Chin's experiment\cite{Chin}. 
The Zak phase\cite{Zak}, which characterizes the winding number of the spin when $k_x$ changes from $-\pi$ to $\pi$, is zero.

In contrast, when the 2-photon process is dominant, the Hamiltonian can then be written as 
\begin{equation}
H=\left((t_s+t_p)\cos(2k_x)/2+\tilde{\Delta}/2\right)\sigma_z-\Omega_2\sin(2k_x) \sigma_y,
\end{equation}
where $\tilde{\Delta}=\Delta-2\hbar\omega$ is the detuning. Such Hamiltonian is equivalent to that obtained in a tilted double-well lattice\cite{Liu}. This Hamiltonian could produce two flat bands in the limit $t_s=t_p=\Omega$ and $\tilde{\Delta}=0$. Moreover, in certain parameter regions, it provides a finite Zak phases $\pm \pi$. As shown in Fig (3) of this supplementary material, when $|\tilde{\Delta}|<\tilde{\Delta}_c=t_s+t_p$, it corresponds to a $2\pi$ rotation of the spin on the $y-z$ plane when $k_x$ changes from $-\pi$ to $\pi$, which leads to a Zak phase of $\pm \pi$ for this pseudo-spin-1/2 system. With changing the value of $\omega$, when $\tilde{\Delta}>\tilde{\Delta}_c$, the contour of spin does not encloses the origin and the Zak phase vanishes.

\end{document}